\documentstyle[aps,prl]{revtex}
\begin{document}
\title{Chaos and Lyapunov exponents in classical and quantal
distribution dynamics}

\author{Arjendu K. Pattanayak and Paul Brumer}

\address{Chemical Physics Theory Group,
\\University of Toronto, Toronto, Canada M5S 3H6}
\date{July 21st, 1997}
\maketitle
\begin{abstract}
We analytically establish the role of a spectrum of Lyapunov exponents in
the evolution of phase-space distributions $\rho(p,q)$. 
Of particular interest is $\lambda_2$, an exponent which quantifies the rate at 
which chaotically evolving distributions acquire structure at increasingly smaller 
scales and which is generally larger than the maximal Lyapunov exponent
$\lambda$ for trajectories.
The approach is trajectory-independent and is therefore applicable to
both classical and quantum mechanics. In the latter case we show that 
the $\hbar\to 0$ limit yields the classical, fully chaotic, result for the 
quantum cat map.  
\end{abstract}
The canonical measure of chaotic Hamiltonian dynamics is the non-zero Lyapunov 
exponent\cite{lyap} which quantifies the rate of exponential divergence of 
neighboring classical trajectories. This measure has resisted translation to the 
Hilbert space of phase space distributions, where the linearity of the 
evolution equation precludes the asymptotic exponential divergence 
of initially ``close" distributions. Rather, chaos in distributions is 
described in terms of the character of the spectrum of the Liouville 
operator\cite{liouv}, a difficult property to access. As a consequence, 
a quantitative diagnostic for the manifestation and characterization of chaos 
in the evolution of distributions has proven elusive, although several conjectures 
and limited diagnostics have been proposed\cite{peres1,peres,ballentine,fox,ap_pb}.

In this article we analytically establish a diagnostic of chaos
for distributions that is independent of trajectories and is valid for 
arbitrary initial distributions and at asymptotic times. 
In particular, we expose the role of an entire spectrum of generalized 
Lyapunov exponents $\lambda_{\gamma}$\cite{schlogl} in the evolution 
of distributions. Although we derive the results by considering phase-space 
trajectories, our final expressions for $\lambda_{\gamma}$ depend solely on 
the evolution of the phase space distribution. We thus have a means of studying 
the role of Lyapunov exponents in the evolution of both classical and quantum 
systems, as shown below.

We also show that the particular exponent $\lambda_2$, which is  generally 
larger than the standard trajectory-based maximum Lyapunov exponent 
$\lambda$\cite{schlogl}, assumes particular importance as the
arbiter of the growth of phase space structure. This growth has been
thought\cite{chirikov} to characterize chaos for distributions. In particular,
we show that (1) a distribution evolves so as to acquire structure at 
increasingly smaller scales at an exponential rate given by $\lambda_2$
(2) the rate at which the information in the system moves to the smaller 
scales (and hence, the rate at which it relaxes) is given by $\lambda_2$ 
and (3) the error in a calculation for a fixed level of resolution increases
exponentially in time with $\lambda_2$.

Consider the usual computation of the Lyapunov exponents\cite{lyap}: 
Let the equations of motion of a point in phase space be $\dot x_i =
f_i(x)$, where $x [\equiv (q,p)$ for Hamiltonian systems] is the vector 
denoting all phase space variables and individual vector components are 
denoted by subscripts. 
Our variables are scaled to be dimensionless so that we may
define a metric in phase space. The equations of motion for the vectors in
the tangent space are obtained by linearizing the equations of motion 
[substituting $x + \varsigma$ into $\dot x_i = f_i(x)$] around this 
fiducial trajectory:
\begin{equation}
\label{eq:Mij}
\frac{d \varsigma_i}{dt}= \sum_j M_{ij}\varsigma_j 
\end{equation}
where $M_{ij}= \frac{\partial f_j}{\partial x_i}$ is the Jacobian matrix at 
the point $x(t)$. If this system is evolved forward in time, then we can 
obtain the maximal Lyapunov exponent as [Theorems A and B, Ref.~\cite{lyap}]
\begin{equation}
\label{eq:theorem}
\lim_{t\to\infty}\frac{1}{t}\ln(|\varsigma(t)|) 
= \lambda(x)
\end{equation}
where $|\cdots|$ defines the standard norm in phase-space and the result
is independent of $\varsigma$ except for a set of vanishing measure.
For a Hamiltonian system the set of exponents is unchanged under the 
transformation $\lambda \to -\lambda$ as a consequence of time-reflection 
symmetry.

We use the foregoing to derive the equations of evolution for the first-order 
derivatives of a distribution. The assumptions necessary in our analysis 
imply that the results are valid for all distributions that evolve 
non-trivially under the Liouville equation and are differentiable: 
This specifically excludes delta-function distributions (point trajectories) 
and time-invariant solutions to Liouville's equation. 
Consider the relationship between the densities at the 
point $x$ and $x + \varsigma$: 
\begin{equation}
\rho(x + \varsigma) = \rho(x) + \sum_i\varsigma_i 
\frac{\partial}{\partial x_i}\rho(x) 
\end{equation}
where we neglect higher-order terms, as in the trajectory-based derivation. 
Taking the total time-derivative of both sides and noting that 
$\frac{d \rho(x)}{dt} \equiv 0 \equiv \frac{d \rho(x + \varsigma)}{dt}$ we get 
\begin{equation} 
\sum_i \frac{d}{dt} \bigg[\varsigma_i(
\frac{\partial\rho}{\partial x_i})\bigg] = 
\sum_i \bigg[ \frac{d \varsigma_i}{dt}
\frac{\partial\rho}{\partial x_i} + \varsigma_i\frac{d}{dt} 
(\frac{\partial\rho}{\partial x_i})\bigg] = 0
\end{equation}
where here and below $\rho$ and its derivatives are evaluated at $x$. 
Using Eq.~(\ref{eq:Mij}), we get 
\begin{eqnarray}
&&\sum_i\bigg[\sum_j M_{ij}\varsigma_j
(\frac{\partial \rho}{\partial x_i})\bigg] + 
\sum_j\bigg[\varsigma_j\frac{d}{dt}
(\frac{\partial \rho}{\partial x_j})\bigg]
= 0\\
&\Rightarrow&
\sum_j\varsigma_j \bigg[ \sum_i \bigg(M_{ij} 
(\frac{\partial\rho}{\partial x_i})\bigg)+ 
\frac{d}{dt} (\frac{\partial\rho}{\partial x_j})\bigg] 
= 0 \\
&\Rightarrow&
\frac{d}{dt} (\frac{\partial \rho}{\partial x_j}) = - 
\sum_i M_{ij}(\frac{\partial \rho}{\partial x_i}),
\label{eight}
\end{eqnarray}
the last equality following from the independence of the various 
$\varsigma_j$. Eq.~(\ref{eq:Mij}) and Eq.~(\ref{eight}) for the evolution 
of $\varsigma$ and $\nabla\rho$ respectively are identical except for a 
minus sign. This makes physical sense: Since the number of points cannot be 
created or destroyed along trajectories, the density sharpens (flattens) 
along the direction in which points move closer (further). This being the case, 
$\lambda$ is directly reflected in the gradient of the distribution 
$\nabla\rho(x(t))$ at asymptotic times at almost every phase space 
point\cite{footnote} in the chaotic region. Remembering the reflection 
symmetry of the Lyapunov exponents, the maximal Lyapunov exponent may hence be 
equally well computed as 
\begin{equation}
\label{eq:lambda}
\lambda(x) = \lim_{t\to\infty}\frac{1}{t}
\ln(|\nabla\rho(x(t))|).
\end{equation}

We now consider the averaged quantites  $\chi_{\gamma}$ where 
\begin{equation}
\chi_{\gamma}\equiv
\bigg (
{\frac{{\rm Tr}[\;|\nabla\rho(x(t))|^{\gamma}]}{4\pi^2 
{\rm Tr}[\rho^{\gamma}(x(t))]}}
\bigg )^{\frac{1}{\gamma}}.
\label{gamma}
\end{equation} 
Here ${\rm Tr}$ indicates the trace or integration over phase space 
and $\gamma$ is an arbitrary real number; note that the denominator
${\rm Tr}[\rho^{\gamma}(x(t))]$ is a constant for Hamiltonian evolution. 
Several interesting properties have been established for similarly constructed 
averages using $\varsigma$ instead of $\nabla\rho$\cite{schlogl}. Using the 
same approach, since Eqs.~(\ref{eq:theorem}) and (\ref{eq:lambda}) are 
identical, we get that the following properties also hold for 
$\lambda_{\gamma}$ defined via Eq.~(\ref{gamma}):
(1) A spectrum of generalized Lyapunov exponents can be defined through 
\begin{equation}
\lim_{t \to \infty} \frac{1}{t}\ln(\chi_{\gamma}) = \lambda_{\gamma},
\label{eq:lambda_g}
\end{equation}
(2) The `usual' maximal Lyapunov exponent $\lambda$ is a member of 
the spectrum: $\lim_{\gamma \to 0} \lambda_{\gamma} = \lambda$ and
(3) $\lambda_{\gamma}$ is concave in $\gamma$; i.e., if 
$\gamma_1 < \gamma_2$, then $\lambda_{\gamma_1} \leq \lambda_{\gamma_2}$.
Note that by defining the averages in terms of $\nabla\rho$ rather than
$\varsigma$, the quantities $\chi_{\gamma}$ and $\lambda_{\gamma}$ are 
independent of trajectories.

We note that the dependence of $\lambda_{\gamma}$ 
on $\gamma$ arises due to the variation in local stretching rates in different
regions of phase space.  Hence, for a system where the local stretching rate 
is a constant, $\lambda_{\gamma}$ is independent of $\gamma$ and equals the
maximum Lyapunov exponent $\lambda$.

Considerable physical insight emerges by considering $\lambda_2$ and the 
quantity $\chi_2(t)$ in Fourier space. That is, consider the Fourier expansion 
of a distribution $\rho(q,p)= \sum_{n,m}\rho_{n,m}(t)f_{n,m}$ where 
$f_{n,m}(p,q)=\exp\{2\pi i(n p +m q)\}$.
We use a two-dimensional discrete Fourier basis for simplicity; all arguments 
generalize to multiple dimensions and Fourier integrals. In Fourier space  
$\chi_2(t)$ is of the form
\begin{equation}
\chi_2^2(t)\equiv \frac{{\rm Tr} [\;|\nabla\rho(x(t))|^2]}
{4\pi^2 {\rm Tr}[\rho^2(x(t))]} =
\frac{\sum_{n,m}(n^2 + m^2)|\rho_{n,m}(t)|^2}{\sum_{n,m}|\rho_{n,m}(t)|^2}.
\end{equation}
Hence ${\chi_2}$ is the root-mean-square Fourier radius, measuring 
the Fourier space extent of the phase space distribution. 
Equation (\ref{eq:lambda_g}) shows that 
\begin{equation}
\lim_{t \to \infty} \frac{1}{t}\ln({\chi_2}) = \lambda_2;
\label{eq:lambda_2}
\end{equation}
or that, for a chaotic system, $\chi_2$ increases exponentially with time.  
Thus, since the higher $(|n|,|m|)$ modes correspond to structure at smaller
scales, a distribution moves exponentially in time, with rate $\lambda_2$,
from structure at a discernible scale to structure at extremely small scales. 
We can relate this to the loss of accuracy associated with the chaotic evolution 
of point trajectories as follows\cite{ford}. 
All the information about the distribution is encoded in the initial Fourier 
basis expansion; as $\chi_2$ grows, this information leaves any finite range 
of $|n|,|m|$ exponentially fast. Attempting to account for 
all Fourier modes corresponds to retaining an infinite amount of information 
and is inconsistent with a finite-resolution measurement in phase-space. Thus, 
Fourier modes with mode-numbers greater than some $n_{\rm max},m_{\rm max}$, 
where $\frac{1}{n_{\rm max}},\frac{1}{m_{\rm max}}$ are the limits of 
resolution in $p,q$ respectively, are effectively non-observable and the 
information therein not retrievable. In this sense the evolution of a chaotic
distribution can be reconciled with the exponential loss of accuracy inherent 
in the chaotic evolution of trajectories, also a consequence of finite 
measurement capability.

Equation~(\ref{eq:lambda_g}) provides deep insights into the relation between 
the Lyapunov exponents and distribution dynamics but should not be regarded as 
a competitive tool to compute Lyapunov exponents when trajectories can be 
calculated.
However, since $\chi_{\gamma}(t)$ no longer relies on trajectories, we may
now study it for quantum mechanical phase-space distributions, thus analyzing 
``quantum chaos", i.e. the effect of chaos on quantal evolution. 
Again, $\chi_2$ proves particularly valuable since, by integration by parts
\begin{equation}
\label{eq:nabla}
\chi_2^2(t)\equiv \frac{{\rm Tr} [\;|\nabla\rho|^2]}
{4\pi^2 {\rm Tr}[\rho^2]} = - \frac{{\rm Tr} [\rho\nabla^2\rho]}
{4\pi^2 {\rm Tr}[\rho^2]}.
\end{equation}
Noting that $\nabla^2\rho \equiv \{p,\{p,\rho\}\} +\{q,\{q,\rho\}\}$, where
$\{\,,\,\}$ denotes the Poisson bracket,  we may now perform the standard 
quantization $\{A,B\} \to -\frac{i}{\hbar}[\hat A,\hat B]$  to yield a quantal 
$\chi_2$ which is independent of the representation.
In the Wigner-Weyl\cite{groot} representation it 
has precisely the classical form [Eq.~(\ref{eq:nabla})], with $\rho$ replaced
by the Wigner function $W(q,p)$.

As an example of the insights afforded by this approach we consider 
the cat map on the torus, which is a classical K-system\cite{arnold} 
and which has been recently shown\cite{ap_pb} to display smooth 
quantum-classical correspondence. The dynamics of this system, whose 
classical and quantal propagators for distributions are known 
analytically\cite{jw}, are those of a kicked oscillator with Hamiltonian\cite{ford} 
\begin{equation}
H=p^2/2\mu +\epsilon q^2/2\sum_{s=-\infty}^{\infty}\delta(s-t/T).
\label{fh}
\end{equation}
restricted to a torus $0\leq q < a$, $0\leq p < b$. The cat map corresponds 
to the choice $\eta=Tb/\mu a =1$ and $\xi=-\epsilon Ta/b=1$  and 
$\alpha=h/ab$ acts as a dimensionless form of Planck's constant for this 
problem. Two other cases of interest we study are the (stable) elliptic map 
($\eta=1, \xi=-1$) and the borderline parabolic case ($\eta=1,\xi=0$). 
We note that this classical system has a constant Jacobian matrix and 
$\chi_{\gamma}\equiv\chi$ is independent of $\gamma$.

We then anticipate that (1) for the stable system $\chi$ oscillates 
as a function of time, (2) $\chi$ grows linearly for the parabolic 
case and (3) a chaotic system has $\chi$ growing exponentially in 
time with the asymptotic rate given by $\lambda$.
This behavior is indeed confirmed computationally, as shown in Fig.~1, for 
these three maps. We emphasize that (1) the behavior is independent of the 
initial distribution; the distributions used here were randomly initialized 
and (2) in the chaotic case, there is no saturation of $\chi(t)$ even 
though the phase-space is bounded.
 
In the quantal case we study the dependence of $\chi_2(t)$ on the degree of
classicality of the system, that is the value of $\alpha$. To eliminate 
kinematical effects of changing $\alpha$ we consider the evolution of Wigner 
distributions which are initially of the Gaussian form
\begin{equation}
W(q,p) = N_\beta \exp [-\frac{(q- \beta q_0)^2}{\beta^2 \sigma_q^2}] 
\exp [-\frac{(p- \beta p_0)^2}{\beta^2 \sigma_p^2}]
\label{beta}
\end{equation}
where $\beta$ is a scaling variable (see below), $N_\beta$ is a normalization 
factor, $(\beta q_0, \beta p_0)$ specifies the location of the Gaussian of 
width $(\beta \sigma_q, \beta \sigma_p)$ in a phase space of dimension 
$\beta^2ab$. Since $\alpha=h/\beta^2ab$, by increasing $\beta$ we approach 
the classical limit while preserving the ratio 
($\beta^2 \sigma_q\sigma_p/\beta^2ab$) of the volume of the initial 
distribution to the volume of phase space. Here we choose this ratio as 
$\approx 0.0175$, sufficiently large to display quantum effects at relatively 
early times. Note that for these large initial distributions, the approximate methods
of measuring the separation of centroids\cite{ap_pb} or the growth of second 
moments\cite{fox} {\em fail} to detect the existence of chaos, even for the 
classical system. Our results are displayed in Fig.~2.

For the classical cat ($h=0,\beta a= \beta b = 1$), shown again for 
comparison, $\chi_2(t)$ grows exponentially after a short transient, with a 
rate equal to $\lambda$. The behavior of $\chi_2$ for the quantized cat map 
is shown for $\alpha=10^{-5}, 10^{-2}$ and $ 10^{-1}$. Near the classical limit 
$(\alpha = 10^{-5}; h=1, \beta a = \beta b = 316.2)$ $\chi_2$ exhibits 
exponential growth, indistinguishable from the classical behavior on the 
finite Fourier grid (of size $2048 \times 2048$) used in our computations. 
However, as $\alpha$ is increased to $10^{-2}~(h=1,\beta 
a= \beta b = 10$) the initial growth of $\chi_2$ is seen to be faster 
than that of the classical cat map, suggesting that the amount of structure 
in the quantal distribution at short times actually {\em increases} relative to 
the classical one. This is indeed the case since one expects the
quantal $W(q,p)$ to adiabatically follow the classical $\rho(q,p)$ but with 
added fringes\cite{kb} (interference structures). Such interference structures 
can be seen in contour 
maps of the evolving distributions. For example, an examination of Figs.~1-3 
of Ref.~\cite{ap_pb} at $t=4T$ shows clearly that the distribution with 
$\alpha =10^{-2}$ has far more structure at finer scales than does the 
distribution with $\alpha=10^{-5}$. This, of course, implies greater support 
for the distribution at larger Fourier node-numbers, leading to larger $\chi_2$ 
values. However, supra-classical growth does not persist for larger $\alpha$ 
or for longer time because a 
second quantal effect kicks in: Quantal distributions resist the growth of 
structure at scales smaller than $\hbar$\cite{kb}, implying that the support of 
the distribution travels slowly across the $\frac{1}{\alpha}$ boundary in 
Fourier space. This is marked by a clear slow-down of the growth at 
$\chi_2 \approx 100 = {\cal O}(\frac{1}{\alpha})$. As $\alpha$ is increased to 
$10^{-1} ~(h=1,\beta a= \beta b = 3.162$), the initial rapid growth saturates
in one time step, resulting in a slowly growing $\chi_2$.
The interplay between the two effects thus implies in general a finite value of 
$\alpha$ at which the quantum system acquires structure maximally
rapidly for early times. Note also that the initial differences between the 
various values of $\chi_2$ in Fig.~2
is exaggerated by the fact that we are using discrete time-steps which are
themselves of the order of $\frac{1}{\lambda_2}$ in our model system.

Since the quantal propagator is a smooth function of $\alpha$, the results at 
larger values of $\alpha$ imply that on the full (infinite) Fourier grid the 
quantal $\chi_2$, even at the smallest values of $\alpha$, will ultimately slow 
down,  reflecting the non-ergodic nature of the quantum map\cite{jw}.
However, and this point is significant, for any finite resolution there is 
always an $\alpha$ sufficiently small such that the classical and quantal 
behavior are indistinguishably chaotic. Thus, the various $\lambda_{\gamma}$ 
govern the behavior of classical and quantal distributions in precisely the 
same manner in the limit $\hbar\to 0$, as manifest in the behavior of $\chi_2$. 
Note that the form of $\chi_2(t)$ enables us to verify this behavior without 
masking effects arising from the choice of broad initial distributions or 
from saturation and relaxation in 
the finite phase-space (see the discussion in Ref.~\cite{ap_pb}). 

In summary, we have provided a quantitative demonstration of the role of 
Lyapunov exponents in the behavior of classical and quantal distributions and 
have identified $\chi_2(t)$ as a measure of the increasingly fine detailed 
structure of the distribution. We have thus related the generalized Lyapunov 
exponents directly to the properties of distributions, independent of 
trajectories. Our approach is valid for arbitrary initial distributions and 
asymptotic times, unlike other methods\cite{fox,ap_pb}, and has been explictly
derived from the definition of Lyapunov exponents, unlike previous 
conjectures\cite{peres1,peres,ballentine}. In particular, it avoids the need to 
introduce perturbations to define the stability of distributions\cite{peres1}
and avoids basis-dependent definitions of chaos in distributions\cite{peres}.
Rather, it provides a consistent understanding of the role of Lyapunov exponents 
in the dynamics of classical distributions and affords a method of connecting Lyapunov 
exponents to the behavior of quantal distributions in the semiclassical limit 
$\hbar\to 0$.

{\bf Acknowledgement:} This research was supported by the Natural
Sciences Engineering and Research Council of Canada. We thank an anonymous
referee for comments emphasizing the existence of a spectrum of Lyapunov 
exponents. AKP would like to thank Salman Habib for useful discussions.

\begin{figure}[htb]
\caption{$\chi$ as a function of time for the classical elliptic, parabolic 
and cat maps.}
\end{figure}
\begin{figure}[htbp]
\caption{$\chi_2(t)$ for the classical ($\alpha=0$) and quantal
($\alpha=10^{-1}$,$10^{-2}$ and $ 10^{-5}$) cat map. }
\end{figure}


\begin{thebibliography}{99}
\bibitem{footnote} We say ``almost every" to exclude unstable periodic 
orbits, a set of zero measure.
\end{thebibliography}
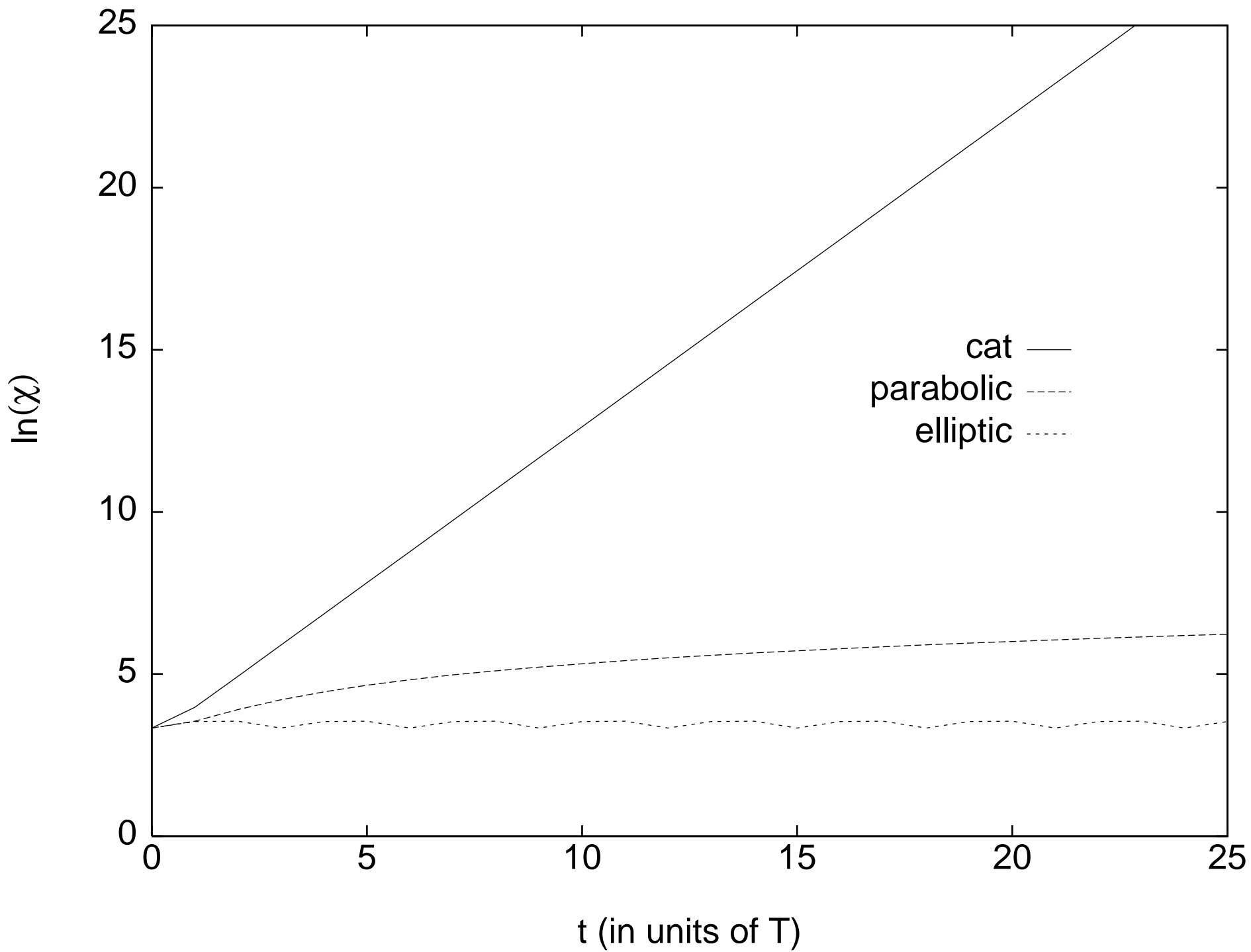
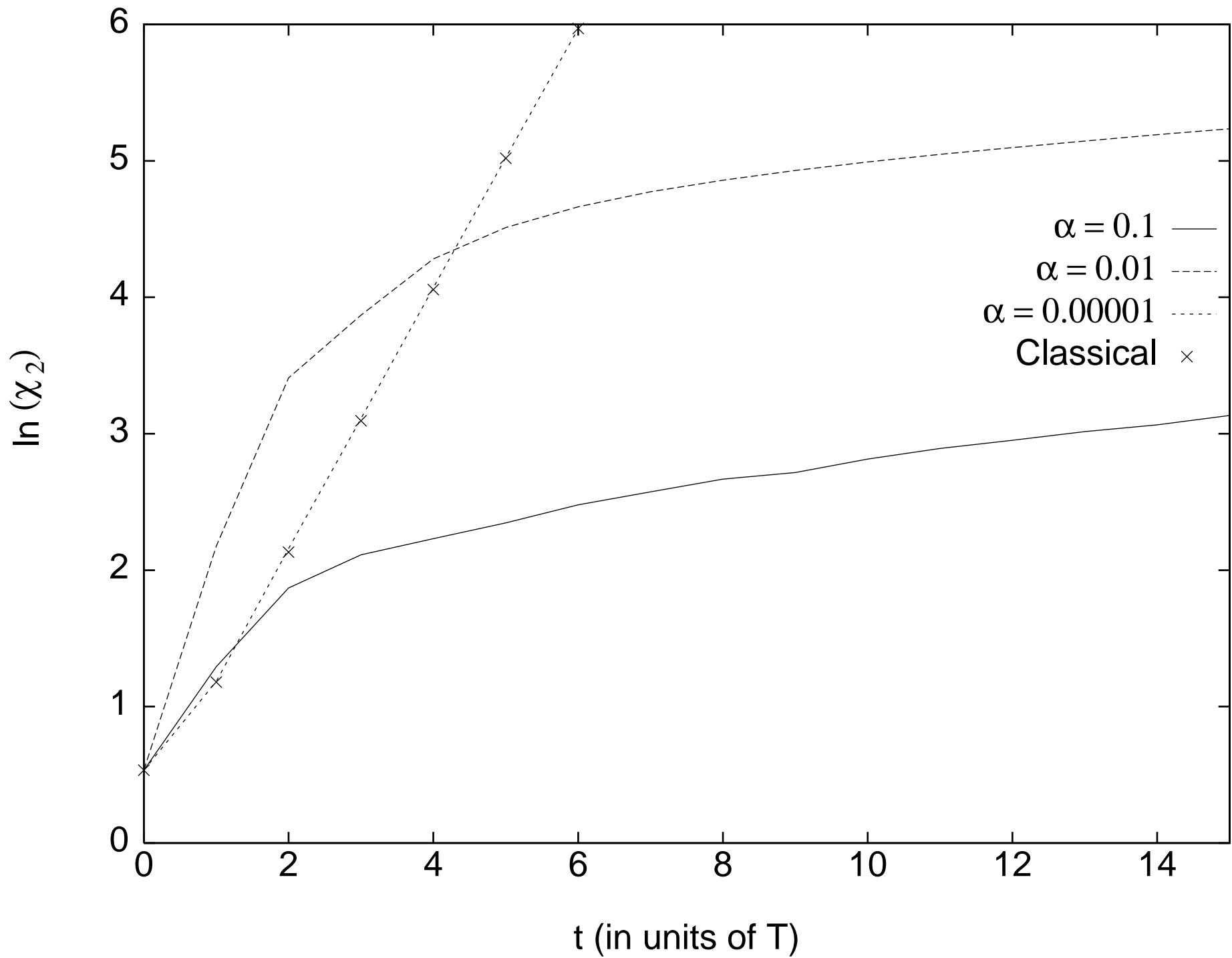
\end{document}